\documentclass[12pt,preprint]{aastex}
\usepackage{lineno}

\newcommand{\kms}{km s$^{-1}$}
\usepackage{graphicx}
\usepackage{epsfig}
\usepackage{times}
\usepackage{natbib}
\usepackage{url}
\usepackage{color}
\usepackage{amssymb,amsmath}
\usepackage{hyperref}
\shorttitle{Fine structures of EUV waves}
\shortauthors{CHANDRA, DEVI, CHEN, JOSHI, \& JOSHI}

\begin{document}

\title{Direct evidence of hybrid nature of EUV waves and the reflection of the fast-mode wave}

\author{Ramesh Chandra\altaffilmark{1}, P. F. Chen\altaffilmark{2,3}, Pooja Devi\altaffilmark{1} }

\affil{$^1$ Department of Physics, DSB Campus, Kumaun University, Nainital -- 263 001, India; \email{rchandra.ntl@gmail.com}}
\affil{$^2$ School of Astronomy \& Space Science, Nanjing University, Nanjing 210 023, China}
\affil{$^3$ Key Laboratary of Modern Astronomy and Astrophysics (Nanjing University),
Ministry of Education, Nanjing 210023, China}
\begin{abstract}
We performed an analysis of the extreme ultraviolet (EUV) wave event on 2022 March 31. The event originated from active region (AR) 12975 located at N13W52 in the field of view of the Atmospheric imaging Assembly (AIA) and exactly at the west limb viewed by the EUV Imager (EUVI) of the Solar Terrestrial Relations Observatory--Ahead (STEREO--A) satellite. The EUV wave was associated with an M9.6 class flare. The event was also well observed by MLSO and COR1 coronagraphs. We revealed here evident coexistence of two components of EUV waves in AIA as well as in EUVI images i.e., a fast–mode wave and a nonwave, which was predicted by the EUV wave hybrid model. The speeds of the fast–mode and non wave EUV wave components in AIA
varies from $\sim$430 to 658 \kms~ and $\sim$157 to 205 \kms, respectively. The computed speeds in STEREO--A for the fast–mode wave and nonwave components are 
 $\sim$520 and $\sim$152 \kms, respectively. Another wave emanated from the source AR and interacted with ambient coronal loops, showing evident reflection in the EUV images above the solar limb. {The speed of the reflected wave in the plane of the sky is $\sim$175 \kms.  }
 %
 With the precise alignments, we found that the fast-mode EUV wave is just ahead of the coronal mass ejection (CME) and the nonwave component is cospatial with the frontal loop of the accompanied CME. The event also showed stationary fronts.


\end{abstract}

\section{Introduction}


There are several large-scale phenomena in the low corona. Among them, filament eruptions, coronal mass ejections (CMEs), and extreme ultraviolet (EUV) waves are commonly observed \citep{ Gopalswamy92, Harrison95, Hudson96, Gopalswamy03, Jing04, Yashiro05, Chen2011, Schmieder12, Warmuth15, Zou19, Chandra22}. EUV waves were originally observed in EUV wavelengths by the EUV Imaging Telescope \citep[EIT;][]{Delaboudiniere95} and then by other telescopes as well \citep{Thompson99, Asai12, Cabezas19, Wang20}. 
Here, we would like to mention that due to their discovery by the EIT telescope, they were historically known as `EIT waves' \citep{Thompson99}. Afterwards, for several reasons, in particular, their controversial nature \citep{Chen16a}, they were denoted by several names, such as EUV waves, propagating bright fronts, solar tsunami, etc.  For consistency, hereafter we will use the name ``EUV waves'' throughout this paper, and use ``fast component'' and ``nonwave component" to distinguish the sub-classes with different natures.

Initially, it was believed that EUV waves are generated by the solar flares-induced pressure pulse \citep{Wu01, Vrsnak02}. Alternatively, EUV waves were proposed to be related to CMEs \citep{Plunkett98, Delannee99, Biesecker02, Chen02}. To investigate the origin of EUV waves, \cite{Biesecker02} studied several EUV waves and found that they are all accompanied by CMEs. \cite{Chen06} selected non-CME associated GOES M- and X-class flares for the years 1997 and 2005 and concluded none of them produced EUV waves. Now it is generally believed that the large-scale EUV waves are associated with CMEs or at least jets \citep{zheng12} rather than the pressure pulse inside the flares.

Kinematics is an important property to identify the nature of waves. If we look into the literature, we found that EUV wave speed varies from a few to more than a thousand km s$^{-1}$, and in some cases, stationary fronts were observed \citep{Delannee99, Chen02, Zhukov09, Chen11, Schrijver11, Asai12, Warmuth15, Chen16a, Long17, Chandra18a, Chandra21}.  Therefore, to explain the speed difference of EUV waves, in particular the existence of subsonic and super-Alfv\'en EUV waves, \cite{Chen02} proposed a hybrid model. According to this model, EUV waves consist of two components: the first is related to fast--mode MHD wave, and the second is the nonwave component related to the stretching of magnetic field lines during the eruption of the flux rope. The nonwave component is actually the CME frontal loop as reported in some of the observations \citep{Attrill07, Chen09}. 
Using the MLSO and EIT observations on 1997 September 9, \citet{Chen09} showed that the EIT wavefronts (we tend to limit the terminology EIT wave  to the nonwave component of EUV waves) are exactly cospatial with the CME frontal loop observed by the MLSO coronagraph. Here, we would like to mention that during the EIT observations, due to the low  cadence, only the nonwave component was  observed in most events and the fast-mode component may have been missed because of its high speed \citep{Chen02, Shibata02, Wills06}. After the launch of the SDO satellite, many studies reported the coexistence of two wave components. These two components are distinguishable in the time-distance plots \citep{Chen02, Chen11, Asai12, Kumar13, Chandra21, Fulara19}. However, the two components can hardly be directly distinguished in the imaging observations since the EUV images are often noisy between the two fronts. Therefore, it would be interesting to find two separate waves in imaging observations.

 Reflection of the EUV wave is one of the interesting phenomena reported in earlier observations, which evidences the presence of wave component of EUV waves. In these reported observations the reflection was observed near coronal holes (CHs) \citep{Gopal09, Dai10, Kienreich12, Zhou22} and ARs \citep[for example,][]{Ofman02,Shen13, Miao19}. 
  The dynamic radio spectra also show type II radio burst reflection, which is the indication of shock wave at coronal heights \citep{Mancuso21}. \citet{Piantschitsch21} analyzed the geometrical properties of EUV wave and CH interaction and described the characterstics of reflected and  transmitted components of waves. It is noticed that the wave reflection was sometimes utilized as strong evidence to support the wave nature of EUV waves, which is logically incomplete. The reflection of the fast-component EUV wave can not rule out the coexistence of a nonwave component of EUV waves, as demonstrated by the time-distance diagram of \citet{Dai10}. Therefore, it would be crucial if we can distinguish the two wave components and the reflection of the fast-mode wave component in imaging observations simultaneously.
  
  The EUV wave event of 2022 March 31 is a nice example because of its following properties: The clear simultaneous  observations of two components of EUV waves in the EUV images as predicated in the hybrid model of \citet{Chen02}, the clear association between the CME frontal loop and the nonwave component, a stationary front at the CH boundary, clean sweep of fast mode EUV wave component when it passes through the CH, and the reflection of EUV wave from the AR above solar disk. To investigate these features of this EUV wave event, the paper is structured as follows: The observational data sets are introduced in  Section \ref{observation}, the results are given in Section \ref{results}, and discussions as well as a summary are presented in Section \ref{sum}.
  
\section{Observational Data Sets}\label{observation}

For this study, we use the data from the following sources:
\begin{enumerate}
\item{{\bf SDO/AIA :} Atmospheric Imaging Assembly \citep[AIA,][]{Lemen12} on board the 
Solar Dynamics Observatory \citep[SDO,][]{Pesnell12} observes the Sun 
in EUV, ultraviolet (UV), and white light wavelengths with high spatial  and temporal resolutions. The pixel size and the cadence of the AIA images are $0\farcs 6$ and 12 sec, respectively. In this study, we use the AIA data in 193 and 171 \AA\ wavelengths.}

\item{{\bf STEREO--A :} Solar TErrestrial RElations Observatory-Ahead  \citep[STEREO--A,][]{Kaiser08} observes the solar full disk in some of the EUV wavelengths  up to 1.7$R_{\odot}$  with an observing angle different from SDO. The pixel and temporal resolution  of the STEREO--A EUVI data in 193 \AA\ are $1\farcs 6$ and 2.5 min, respectively. Together with the EUVI telescope, the STEREO--A COR1 \citep{Thompson08} observes the associated CME from 1.5 to 4$R_{\odot}$.}

\item{{\bf MLSO:} 
The white-light Mark-III K-Coronameter (MK3) has been operating at the Mauna Loa Solar Observatory (MLSO)  since 1980 \citep{Mac83}. This coronagraph observes the scattered radiation of photospheric brightness, which is scattered by free electrons in the lower corona. The field-of-view (FOV) of this coronagraph is from  1.12 to 2.45 $R_{\odot}$. The pixel size and the cadence of this instrument 
are 20\arcsec\ and 3 min, respectively.
}

\item{{\bf LASCO :} The Large Angle and Spectrometric Coronagraph 
\citep[LASCO;][]{Brueckner95} comprises of two coronagraphs, namely, C2 
and C3. These coronagraphs observe the Thomson-scattered visible light through a broadband filter. The FOVs of C2 and C3 are  1.5--6$R_{\odot}$ and  3.7--30$R_{\odot}$,  respectively.}

\end{enumerate}

\begin{figure*}[h]
\centering
\includegraphics[width=0.7\textwidth]{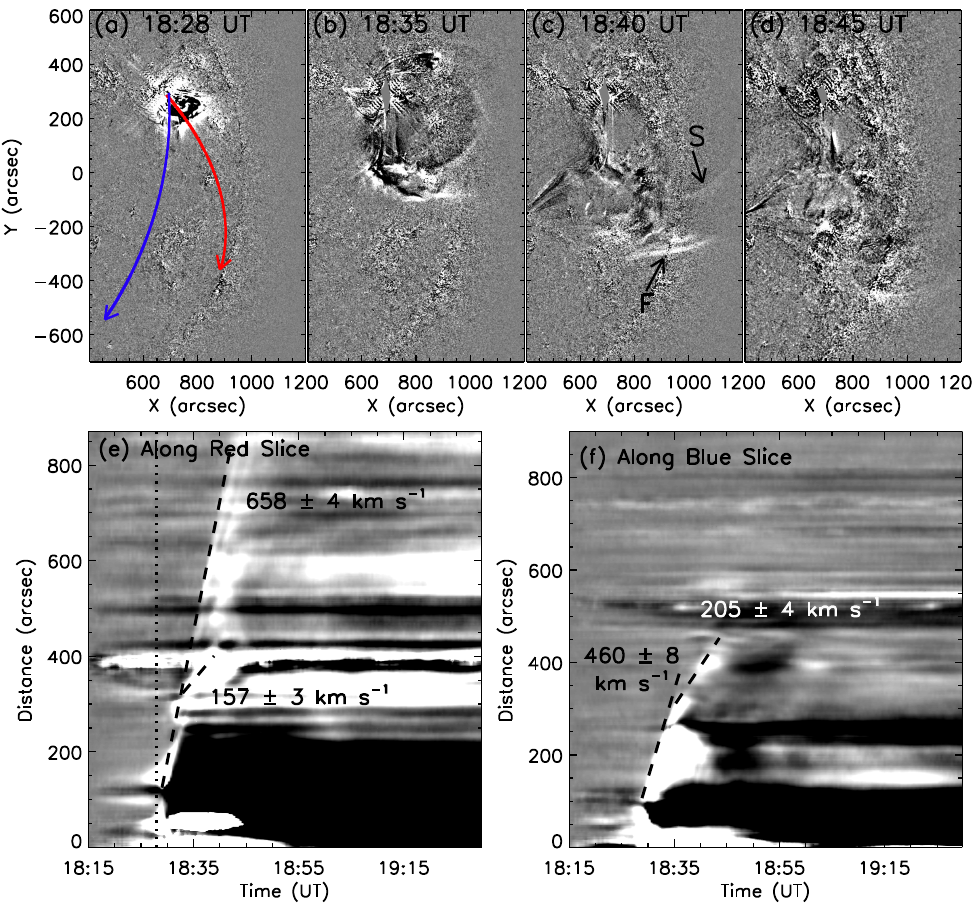}
\caption{ Panels (a--d): Evolution of the EUV waves in AIA 193 \AA\ running difference images. Panels (e--f): Time-distance plot along the red and blue slices drawn in panel (a). The fast-mode and nonwave components of the EUV wave are labeled as `F' and `S' in panel (c). An animation of this figure is available. The animation starts at 18:15 UT and ends at 19:15 UT. The real-time duration of the animation is 57 s.  The black vertical dotted line in panel (e) shows the start time of the wave, i.e., $\sim$18:28 UT.}

\label{fig1}
\end{figure*}

\section{Results}\label{results}

\subsection{Active region overview}
The NOAA AR 12975 appears at the east limb on 2022 March 22 as a $\beta$ magnetic configuration. As the AR  evolves with time, it turns into the $\beta\gamma$ configuration on 2022 March 29, and the $\beta\gamma\delta$ 
configuration on 2022 March 30. Finally, on 2022 April 04, the AR goes over the west limb with the less complex $\beta\delta$ magnetic configuration. 

The flare activity in the AR starts just before it turns into the $\beta\gamma$ configuration on 2022 March 28 and it continues up to 2022 April 02. During its disk passage, the region produces 51 C-class, 10 M-class, and one X-class flares.  For our study, we choose the EUV wave event on 2022 March 31 associated with the GOES M9.6 flare because of its peculiar nature. 
The AR is located at N13W52 on that day. According to GOES observations, the flare starts  at 18:17 UT, peaks at 18:35 UT, and nearly disappears after 18:45 UT. The flare is impulsive and short-lived in nature and is associated with a slow halo CME which  first appears at 19:12 UT in the LASCO C2 FOV. The speed of the observed CME is 489 \kms\ with an acceleration of 0.12 m s$^{-2}$. 

\subsection{Two components of EUV wave}  

Immediately after the flare onset, an EUV wave is visible on the solar surface and above the limb almost in all AIA EUV wavelengths. We create the running difference images (by subtracting the image one minute earlier) in AIA 193 \AA\ to examine the evolution of the EUV wave, which is displayed in panels (a--d) of Figure \ref{fig1}. For a detailed view of the wave development, we also attach herewith the running difference animation. Initially, the wave is visible in all directions as a circular shape and shows a single component. 
As time advances, two components become clear at $\sim$18:29 UT. We label these two components as `F' and `S' in Figure \ref{fig1}(c). These two components are visible in the south direction of the source region. It is clear from the images and accompanying movie that the distance between these two components increases continuously. This confirms that the two components are the fast-mode and the nonwave components of the wave.

To quantitatively study the kinematics of EUV wave, we select  two slices from the source region to the south direction as indicated by the red and blue arrows in Figure \ref{fig1}(a), and the  corresponding time-distance diagrams in AIA 193 \AA\ are plotted in Figures \ref{fig1}(e) and \ref{fig1}(f), respectively.
In Figure \ref{fig1}e, a fast-moving wave is clearly discernible across all the distance, and a slowly-moving wavelike feature branches out at a distance of 300$\arcsec$.
{It is interesting to note  that the fast-mode wave starts $\sim$ 5 min earlier than the nonwave component.}
By fitting the two propagating features with straight lines, we get their propagation speeds. The calculated speeds for the fast-mode and the nonwave components are found to be $\sim$658$\pm$4 \kms and 157$\pm$3 \kms, respectively. From the time-distance analysis, we can see that the fast--mode wave travels a long distance up to 850$\arcsec$ from the starting point of the slice. When it arrives at a distance of 600$\arcsec$, a stationary front is produced.
For comparison, in Figure \ref{fig1}(f) we plot the time-distance diagram along the blue slice, which runs across some transequatorial loops. In this panel, we can see not only the fast-mode and nonwave components of the EUV wave with speeds of $\sim$460$\pm$8 \kms and 205$\pm$4 \kms, respectively, but also two stationary fronts at distances of 150$\arcsec$ and 550$\arcsec$, respectively. The formation the stationary fronts is presumably due to wave mode conversion.

\begin{figure*}[t]
\centering
\includegraphics[width=0.9\textwidth]{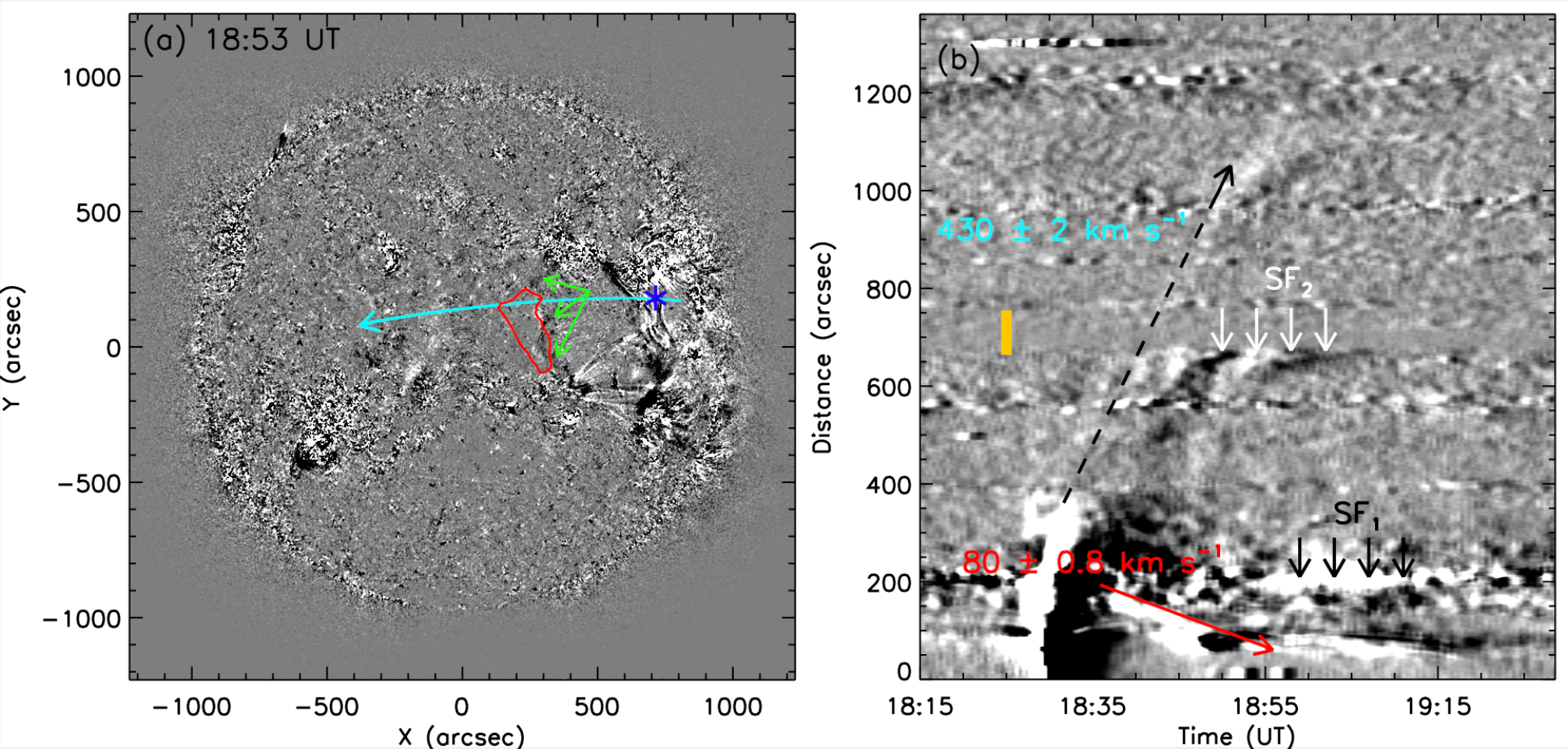}
\caption{Left: AIA 193 \AA\ difference image with the location of the CH (shown by red contour), stationary front, SF$_2$ (green arrows), location of stationary front, SF$_1$ (blue asterisk), and the location of the slice (cyan arrow). Right: Time-distance plot with fast--mode EUV wave (cyan arrow), slow mode wave (red arrow), and the location of SF$_1$ (white arrows) and SF$_2$ (black arrows). The yellow vertical thick bar represents the location of CH. }
\label{fig2}
\end{figure*}

\begin{figure*}[h]
\centering
\includegraphics[width=0.7\textwidth]{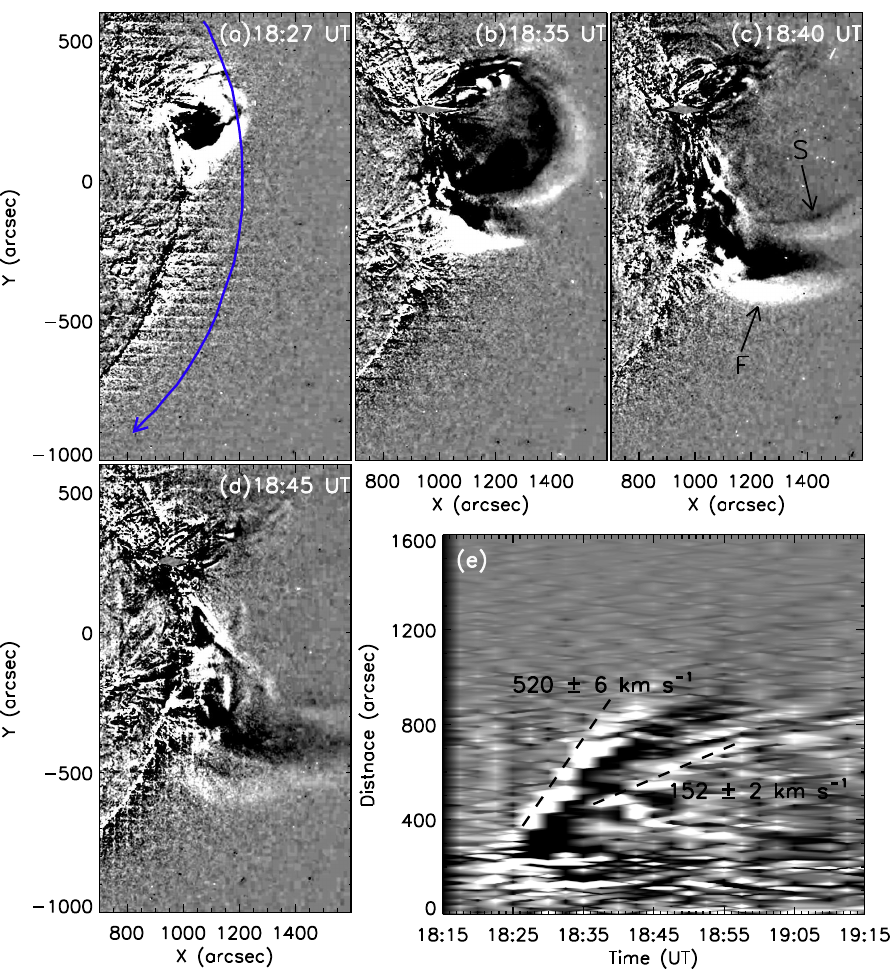}
\caption{Panels (a--d): Development of EUV wave in STEREO--A 195 \AA\ running difference images. Panel (e): Time-distance plot along the blue slice drawn in panel (a). The fast-mode and nonwave components of the EUV wave are labeled by `F' and `S', respectively, in panel (c).}
\label{fig3}
\end{figure*}

The propagation of the EUV wave in the AIA 193 \AA\ wavelength toward the east direction is shown by the time-distance analysis in Figure \ref{fig2}. The location of the slice is shown in Figure \ref{fig2}(a). 
Note that, the small transient hole in Figure \ref{fig2}(a) above the cyan slice corresponds to the interface between the active region AR12976 and the host active region AR12975. Some magnetic field lines overlay AR12976 and others overlay AR12975. The formation of the transient hole is due to the stretching up of the closed field lines in AR12975 during the CME eruption. The time-distance plot corresponding to the slice in Figure \ref{fig2}(a) is given in Figure \ref{fig2}(b).
This time-distance plot reveals the following features: a fast--mode EUV wave having a speed 430$\pm$2 km s$^{-1}$ and two stationary fronts SF$_1$ and SF$_2$. The locations of SF$_1$ and SF$_2$ are at $\sim$200$\arcsec$ and 650$\arcsec$ from the origin site, respectively. The fast-mode wave propagating in the east direction first hits the footpoint of the coronal loop. As a result of this interaction, a slowly-moving wave is excited, which travels toward the west direction with a speed of 80$\pm$0.8 km s$^{-1}$. 
The extremely slow speed of 80 s$^{-1}$ is probably due to the projection effect, where a reflected fast-mode wave propagates mostly down to the solar surface.
Part of the fast--mode wave is converted to a slow--mode wave and is observed as SF$_1$. After this first interaction, the fast--mode wave becomes fainter. At a distance of 650$\arcsec$, it interacts with a CH. This interaction again produces another stationary front, i.e., SF$_2$ at the boundary of CH, whilst this fast--mode wave continues to cross the CH successfully.
From the figure, we see some alternating bright and dark features in SF$_2$. These structures may be seen due to that there are quasi-periodic fast-mode propagating fronts associated with the CME eruption, not just a single coronal Moreton wave \citep[see for example,][]{Liu12, Shen19, Sun22}. Each front of the wave train experiences mode conversion, forming quasi-periodic slow-mode waves as indicated by the alternating bright and dark features in our Figure \ref{fig2}(b).

In addition to the AIA view, the wave was also observed with the STEREO--A satellite from another viewing angle, from which the eruption is a limb event. The evolution of the wave in STEREO--A 195 \AA\ is shown in Figures \ref{fig3}(a--d). Similar to the AIA observations, the wave is visible as a single wave with a circular shape in its initial stage. Later on, the wave is mainly visible in the west and south directions, where the wave is mostly above the limb. As time progresses, we find that the wave splits into two components shown by the arrows and labeled as `F' and `S' in Figure \ref{fig3}(c). The distance between `F' and `S' components increases continuously.

To examine the nature of these two components, we compute their speeds. For this purpose, we select a slice towards the southwest direction as the wave is more prominent in this direction. The position of the selected slice is shown by the blue arrow in Figure \ref{fig3}(a) and the time-distance plot along this slice is displayed in Figure \ref{fig3}(e). The computed speeds for the `F' and `S' components are $\sim$520 $\pm$ 6 km s$^{-1}$ and $\sim$152 $\pm$ 2 km s$^{-1}$, respectively. The values of their speeds allow us to conclude that the `F' component is a fast-mode MHD wave and the `S' component is the nonwave component as predicted in the model of \cite{Chen02}. 
It can also be seen that the fast-mode wave `F' starts slightly earlier than the nonwave component `S'. 
Moreover, many forward propagating features with small slopes come from source region, and they are probably erupting structures inside the CME frontal loop (according to the magnetic field line stretching model, the slower component EUV wave corresponds to the CME frontal loop).
The interesting point of these observations is the clear visibility of the fast-mode and the nonwave components of EUV waves in images as well as in time-distance plots.

\begin{figure*}[t]
\centering
\includegraphics[width=0.6\textwidth]{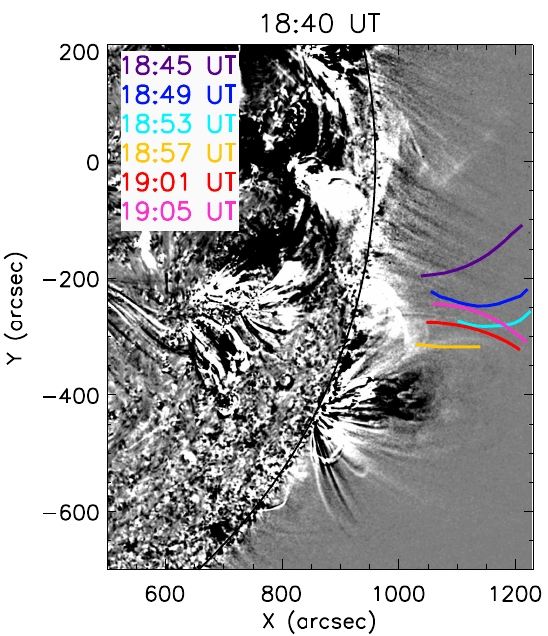}
\caption{Reflection of EUV wave through AR located at the southwest limb in AIA 171 \AA\ wavelength.
{The incident and reflected wavefronts are drawn with purple/blue/cyan and yellow/red/pink colours, respectively.}
An animation is attached with this figure, which starts at 18:17 UT and end at 19:15 UT. The real-time duration of the animation is 12 s. }
\label{fig4}
\end{figure*}

 \begin{figure*}[t]
\centering
\includegraphics[width=0.7\textwidth]{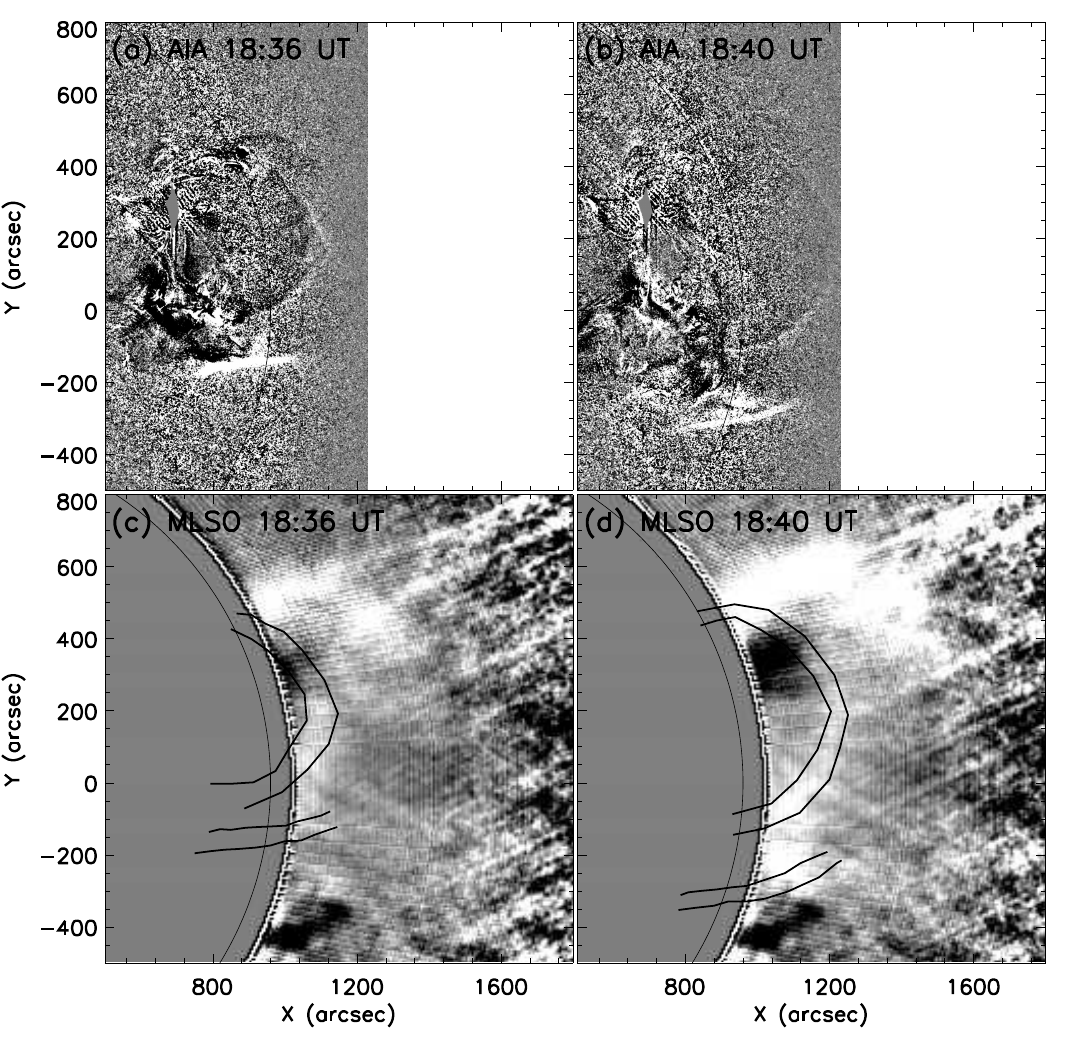}
\caption{Two snapshots of the EUV wave evolution with fast and nonwave components in AIA 193 \AA\ (a, b panel) and images of the CME observed by MLSO coronagraph (c, d panel). The contours of the fast and nonwave component of the EUV wave observed in AIA 193 \AA\ are overplotted in MLSO CME images by black contours.}
\label{fig5}
\end{figure*}

\subsection{Reflection of EUV Wave}

 As we mentioned in the introduction section, the fast--mode EUV wave can be reflected at the boundary of CHs as well as ARs. This is a clear signature of the presence of the real wave component of EUV waves. In the literature, the EUV wave reflection was mainly found during its passage on the solar disk, and the reflection above the limb has never been reported.

 In the present observation, the southward propagating part of the EUV wave encounters another AR NOAA 12972 at $\sim$18:42 UT, which is located at solar X =900$\arcsec$ and solar Y= -400$\arcsec$. After their interaction, part of the wave is reflected and part of it penetrates through the AR. 
 Due to the AR location at the limb, its coronal loops are above the limb, so the incident wave and the reflected wave are seen above the limb. The reflection of an EUV wave above the solar limb is one of the newly reported characteristics of this event. To visualize the reflection and the transmission of the wave, the propagation of the EUV wave toward the south direction along with the incident and reflected wavefronts at different times are displayed in Figure \ref{fig4} with different colors. We also refer the reader to the accompanying movie related to this figure. {The full disk  movie of the event is  available at \url{https://www.lmsal.com/nitta/movies/aia_cadence/aia_0171_pdiffb_0144_sum_2
0220331_1810/AIA_0171_PDIFFB_0144_SUM_20
220331_1810_j.html.} }
 The reflected wave is visible at 18:57 UT and it continues to be seen until 19:04 UT. The transmitted wave continues to move to the south direction and keeps noticeable until 18:53 UT. 
{We measure the speed of the reflected wave in the plane of the sky, which is found to be $\sim$175 \kms.}
These values are less than the speed of the incident wave, which confirms the finding of \citet{Gopal09}. However, the reflection of the wave is not visible in the STEREO--A data set, which could be due to its different viewing angle and the low cadence.

\subsection{Connection between EUV waves and the CME}

Several studies have been performed on the association between EUV waves and  CMEs. However, this association is still debatable. In this section, we compare the spatial link between the different components of the EUV waves and the accompanied CME. The MLSO coronagraph provides the observations of the CME in the inner corona from 1.12 to 2.45$R_{\odot}$. Since the location of the EUV wave source site is close to the west limb in the AIA data sets, such a location provides an excellent opportunity to explore the connection between the EUV waves and the CME. Figures \ref{fig5}(a) and \ref{fig5}(b) display two snapshots of the AIA 193 \AA\ showing the two wave components. In the `c' and `d' panels of the figure, we coalign these EUV images  with the MLSO coronagraph images. 
The contours of the EUV wave are drawn by black contours overlaid on the MLSO images. This coalignment evidences that the fast-mode component of the EUV waves is ahead of the CME frontal loop, and the nonwave component of the EUV waves is well cospatial with the CME frontal loop. 
{ It is seen that fast-mode waves are visible in the south and east, but not in the north. The reason is that the eruption happened near the southern edge of the source active region, hence the magnetic field is much stronger in the north, and fast-mode waves tend to be refracted toward weak magnetic field, avoiding the strong magnetic field in the north. Besides, the eruption was inclined toward south (also influenced by the strong magnetic field in the north), favoring the detection of the fast-mode wave in the south.}

In the model of \cite{Chen02}, the CME-driven shock exactly corresponds to the fast-wave component of EUV waves, as also confirmed by various literature \citep[e.g.,][]{fras19, feng20}. It has been established that the EUV wave fronts are dome-shaped structure, not only limited at low altitudes (though they are brighter at low altitudes), e.g., in the simulation of \citet{Chen02} and in the observation of \citet{vero10}.
Note that the bright regions in the north and the south of the coronagraph are the helmet streamers present in the background, which become clearly visible in the LASCO C2 and C3 observations. Note that the evolution of the CME is available at LASCO catalog (\url {https://cdaw.gsfc.nasa.gov/CME_list/UNIVERSAL/2022_03/jsmovies/2022_03/}). 

\begin{figure*}[h]
\centering
\includegraphics[width=0.7\textwidth]{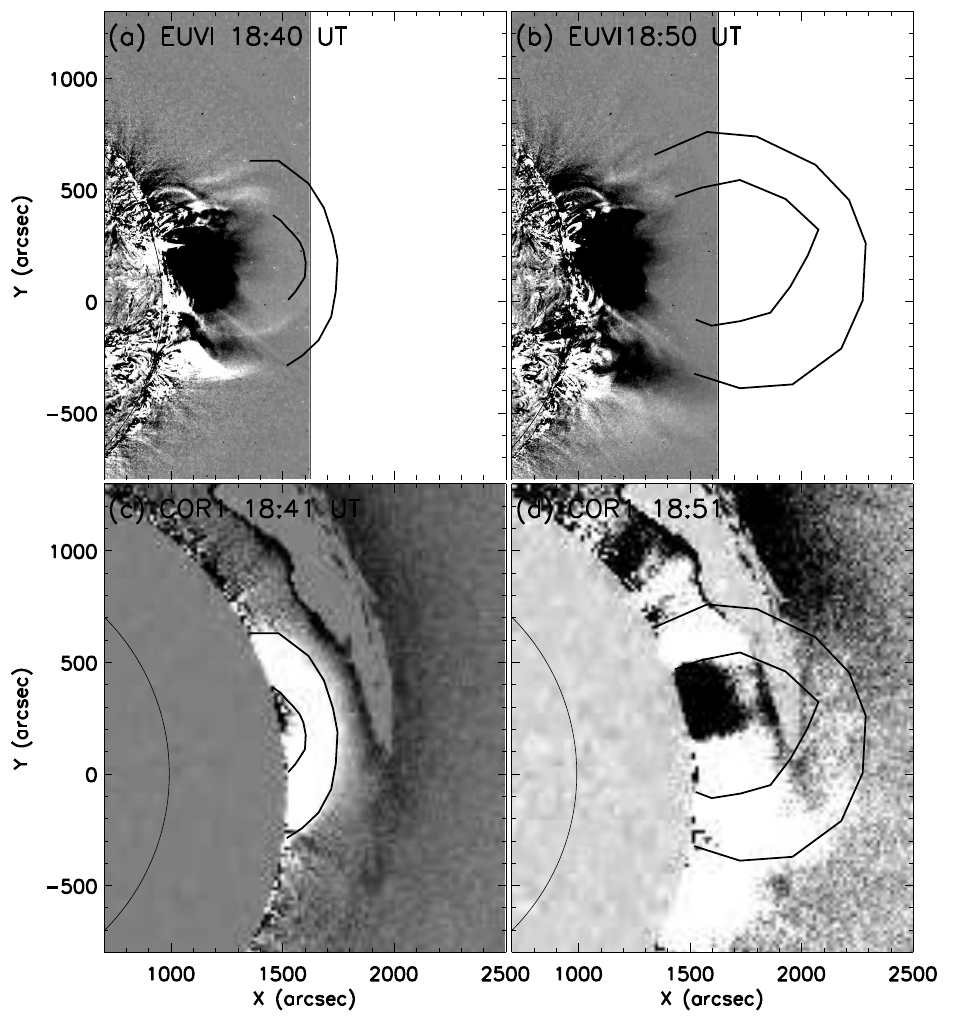}
\caption{ Two snapshots of the EUV wave in STEREO-A EUVI 195 \AA\ (a, b panel) and images of the CME observed by COR1 (c, d panel). The COR1 CME contours are overplotted by black curves in STEREO-A EUVI as well as in COR1 images.}
\label{fig6}
\end{figure*}

The two wave components are also well observed in the STEREO--A 195 \AA\ images at the west limb. To see the spatial relationship between the fast and nonwave EUV wave components from another view angle, we overplot the STEREO--A 195 \AA\ image  on the COR1 CME data. This comparison  also confirms that the 
fast--mode component is ahead of the CME frontal loop. The alignment is presented in Figure \ref{fig6}. The black contours represent the outline of the CME seen in the COR1 coronagraph. This confirms the results derived from the AIA and MLSO data sets.


\section{Discussions and Summary}\label{sum}

In this paper, we presented the two-viewpoint observations of the EUV wave event on 2022 March 31 using the SDO/AIA, STEREO--A/EUVI data, and the coronagraph data of MLSO and COR1. The EUV wave event was associated with a GOES M9.6-class flare along with a halo CME. Our main results are summarized as follows:

\begin{itemize}
    
    \item [--]{Two components of the EUV wave are clearly and simultaneously visible in the AIA 193 \AA\ and in the STEREO--A 195 \AA\ images as well as the time-distance diagrams. {From the time-dsitance diagram, it is observed that the fast-mode wave component starts $\sim$5 min earlier than the slow component, i.e., the nonwave component.}
    }

    \item [--]{We observed two stationary fronts, namely, SF$_1$  and SF$_2$, respectively, when the fast--mode wave encounters the boundary of the source AR and a remote CH.}

    \item[--] {Reflection of the fast-component EUV wave was observed when the wave hit an AR 12972 located in the south direction (about 650\arcsec\ from the wave origin site).}

    \item [--]{The fast and nonwave components of EUV waves are located ahead of the leading edge and at the frontal loop of the CME, respectively. The fast--mode components of the EUV waves observed by the STEREO--A and COR1 instruments are consistent with the result of the AIA and MLSO observations. }
\end{itemize}

These results might shed light to the understanding of various types of EUV waves. To resolve the controversial features of EUV waves, a hybrid model was proposed by \citet{Chen02}. According to their model, EUV waves have two components, a fast component which is a real fast-mode MHD wave or shock wave, and a slow component, which is a nonwave generated by the successive stretching of the magnetic loops overlying the erupting flux rope. Later on, the analysis of many events demonstrated the existence of these two components in various  observations \citep{Chen11, Asai12, Chandra16, Chandra21}. 
In most cases the two components are clear only in the time-distance plots of the wave propagation \citep{Chen09, Chandra18a, Chandra21}, and the simultaneous observations of two distinct components were rarely reported in imaging observations directly. The identification of each component is very important especially in the study of solar energetic particle events (SEPs) and  the type II radio bursts association with EUV waves. Our analysis revealed the clear evidence of fast--mode EUV wave component and the nonwave component not only in the time-distance diagrams, but also in the AIA and STEREO--A images. 

{In our time-distance analysis, the fast-mode component of EUV wave starts $\sim$ 5 min before the slow-component.  Such a scenario was often interpreted as “super-expansion of CMEs”. However, in the hybrid model, it is a natural result: When the coronal magnetic field lines are concentric semicircles, the slow-component is $\sim$ 3 times slower than the fast-mode wave; When the coronal magnetic field lines are stretched in the vertical direction, the slow component would be much slower, more than 3 times slower; When the coronal magnetic field lines are stretched in the horizontal direction, the slow component would approach the fast-mode wave (and may even faster than the fast-mode wave when the field lines are extremely oblate in the horizontal direction) since the slow component is an apparent motion in the hybrid model. In the case of the event under study, it is probably that the coronal magnetic field lines near the source region are stretched in the horizontal direction so that the ''slow-component" is indistinguishable from the fast-mode wave near the boundary of the source region.}
The presence of a stationary front in the time-distance diagram towards south direction ($\sim$600$\arcsec$ away from the wave origin location) could be due to the  mode conversion as proposed in \citet{Chen16b}.
 
The encounter of the EUV waves with CHs was reported earlier \citep{Gopal09, Dai10, Kienreich12}. In those observations, it was reported that the fast--mode EUV wave can be reflected at the boundary of the CHs, which corresponds to a magnetic separatrix. There are also studies, where it was reported that a CH does not affect the fast--mode EUV wave \citep{Chandra22}. This indicates that it is not the magnetic separatrix that reflects the fast--mode waves, and it is the strong gradient of magnetic field across the separatrix that leads to the reflection. In our current observations, we found that, when the fast mode EUV wave encounters a CH, part of it converts into a stationary front labeled as SF$_2$ and the remaining part crosses the CH successfully. The formation of the stationary front is most probably due to the rapid change of magnetic connectivity at the  CH boundary, i.e., a magnetic separatrix, where the fast--mode wave  partially converts into a slow-mode wave, hence forming the stationary fronts as proposed by \citet{Chen16b}. In addition to the SF$_2$ stationary front, we have also observed the stationary front SF$_1$, which could be due to the same reason since it is located at the boundary of the source AR, which also corresponds to a magnetic separatrix. 

Another interesting phenomenon in this event is the reflection of the wave through  the AR 12972, which is located about 650\arcsec\ away from its origin site. The reflection of EUV waves was reported in earlier observations when they pass through CHs on the solar disk. \citet{Miao19} reported the reflection and the refraction through the AR on solar disk for the event of 2011 March 10. The cases of total internal reflection from CH was also reported by \citet{Zhou22}. Furthermore, dynamic radio spectra also indicate the reversal of the type II radio bursts, which confirms the shock--wave reflection. However, there were no imaging observations of EUV wave reflection above the solar limb. Therefore, the present event provides the direct evidence of wave reflection above the limb. 

The EUV waves are associated with a partial halo CME observed by the MLSO, LASCO, and STEREO COR1 coronagraphs. There were studies on the association between the EUV waves and CMEs. However, it is not clear how the two components of EUV waves match the accompanied CME. The present study sheds light on this coupling. Our analysis shows that the fast--mode component of EUV wave is ahead of the CME, which supports the earlier studies \citep{Chen06, Kwon14, Kwon17, Devi22} and the  nonwave component is cospatial with the CME frontal loop. This confirms the study done by \cite{Chen06}. 

It is noted that each single aspect of such results was reported earlier but never reported in a single event, which is a peculiarity of this event. Recent numerical simulations further verified such a global picture \citep{Guo2023}. \\

{\bf Acknowledgments}\\
 {The authors are grateful to the referee for the comments and suggestions. We} thank the open data policy of the SDO and STEREO, and SOHO teams.  P.F.C is financially supported by the National Key Research and Development Program of China (2020YFC2201200) and NSFC (12127901). P.D. is supported by CSIR, New Delhi. 
\bibliographystyle{apj}
\bibliography{reference}


\end{document}